\documentclass[aps,prb,twocolumn,showpacs,superscriptaddress]{revtex4-1}
\usepackage{amssymb,amsmath}
\usepackage{graphicx}
\usepackage{dcolumn}
\usepackage{bm}
\usepackage{color}

\begin{document}

\title{In the search of new electrocaloric materials: Fast ion conductors}
\author{Claudio Cazorla}
\email{c.cazorla@unsw.edu.au}
\affiliation{School of Materials Science and Engineering, UNSW Australia, Sydney NSW 2052, Australia \\
Integrated Materials Design Centre, UNSW Australia, Sydney NSW 2052, Australia}
\email{c.cazorla@unsw.edu.au}

\begin{abstract}
We analyse the effects of applying an electric field on the critical temperature, $T_{s}$,
at which superionicity appears in archetypal fast ion conductor CaF$_{2}$ by means of 
molecular dynamics simulations. We find that the onset of superionicity can be reduced 
by about $100$~K when relatively small electric fields of $\sim 50$~KV~cm$^{-1}$ are employed. 
Under large enough electric fields, however, ionic conductivity is depleted. The normal to 
superionic phase transition is characterised by a large increase of entropy, thereby 
sizeable electrocaloric effects can be realised in fast ion conductors that are 
promising for solid-state cooling applications.
\end{abstract}
\maketitle

$\emph{Introduction.}$~In electrocaloric materials, the adiabatic switch of an electric field causes a change in 
the temperature of the system that is equal to 
\begin{equation}
\Delta T = - \int_{0}^{\epsilon} \frac{T}{C_{p}} \cdot \left( \frac{\partial S}{\partial E} \right)_{T}~dE~,
\label{eq:deltat}
\end{equation}
where $C_{p}$ represents the heat capacity, $S$ the entropy, and $E$ the external electric
field. Ferroelectrics are the archetypal electrocaloric compounds:~\cite{scott11} They 
exhibit an spontaneous electrical polarization below a certain critical temperature, $T_{C}$, 
that can be shifted by an external electric field. Electrocaloric effects are observed at 
$T>T_{C}$, when the system is paraelectric but responds to the presence 
of an external electric field by aligning its dipole moments parallel. In those conditions, 
the adiabatic switch of an electric field causes a positive $\Delta T$ in the crystal because 
the entropy in the ordered state is smaller than in the paraelectric phase 
[$\Delta S < 0$ hence $\Delta T > 0$, see Eq.~(\ref{eq:deltat})]. Conversely, when the electric 
field is adiabatically removed the material gets cooler ($\Delta S > 0$ hence $\Delta T < 0$). 
Several thermal cycles based on the electrocaloric effect just described can 
be engineered, which are promising for solid-state refrigeration applications.~\cite{scott11} 
Nevertheless, ferroelectric materials present a series of technical issues, like 
for instance the presence of ferroelectric domains and leakage currents, that are 
hindering the development of practical electrocaloric applications.~\cite{scott11} 
Therefore, is very desirable to find new electrocaloric materials with improved 
cyclability and electrical resistivity features. Promising materials rivaling ferroelectrics, 
overall, must display a large change of entropy upon the application of an external electric 
field.    

\begin{figure}[t]
\centerline
        {\includegraphics[width=1.0\linewidth]{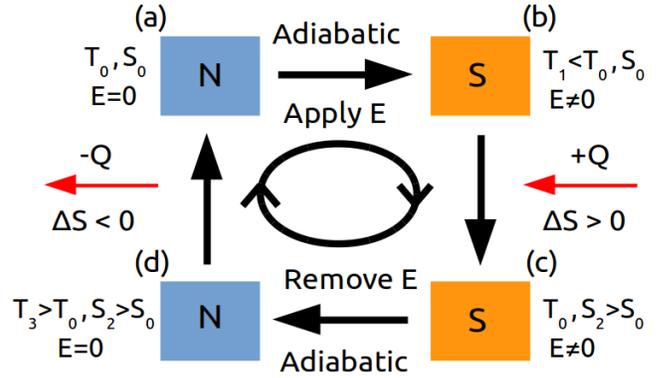}}
\vspace{-0.25cm}
\caption{Schematic diagram of an electrocaloric cooling cycle involving FIC at 
$T < T_{s}$. ``N'' and ``S'' represent the normal and superionic states. 
(a)~$\to$~(b)~An electric field is adiabatically applied that triggers superionicity, 
hence the crystal gets cooler. (b)~$\to$~(c)~The system receives heat from the environment, 
hence $T$ and $S$ increase. (c)~$\to$~(d)~The electric field is adiabatically removed, hence 
the system becomes normal and $T$ increases. (d)~$\to$~(a)~Heat is ejected from the system.}  
\label{fig:intro}
\end{figure}

Long time ago, Kharkats \emph{et al.} proposed, based entirely on theoretical arguments, that 
the application of an external electric field could drastically lower the critical temperature, 
$T_{s}$, at which superionicity appears in fast ion conductors (FIC).~\cite{kharkats} Superionicity 
refers to the unusually large mobility (of the order of $1$~$\Omega^{-1}$~cm$^{-1}$) that a particular 
atomic species in a multicomponent ionic material acquires when temperature is raised (below its 
melting point). The normal to superionic phase transition is experimentally characterised by a 
sudden increase in the heat capacity and entropy of the crystal ($\Delta S 
\sim 10$~J~mol$^{-1}$K$^{-1}$).~\cite{madden04}    
The possibility of externally estimulating superionicity by means of an electric field has
been, to the best of our knowledge, totally overlooked to date. This idea, however, 
has the potential to motivate original searches on new electrocaloric materials for
solid-state cooling applications, as we schematically show in Fig.~\ref{fig:intro}. 
In this molecular dynamics work, we show that $T_{s}$ can be modified appreciably in 
archetypal fast ion conductor CaF$_{2}$ by applying an external electric bias. 
In particular, we show that when $E$ is relatively small ($\sim 10$~KV~cm$^{-1}$) $T_{s}$ can be shifted 
down by about $100$~K. However, under the action of larger electric fields ionic conductivity is depleted. 

$\emph{Methods.}$~Our $(N, P, T)$ simulations were performed with the LAMMPS code,~[\onlinecite{lammps}] 
keeping the temperature and pressure fluctuating around a set-point value by using Nose-Hoover 
thermostats and barostats. Large boxes containing $6,144$ atoms were simulated over long times 
of $\sim 100$~ps, and periodic boundary conditions were applied along the three Cartesian 
directions. Newton's equations of motion were integrated using the customary Verlet's 
algorithm and a time-step length of $10^{-3}$~ps. A particle-particle particle-mesh 
$k$-space solver was used to compute long-range van der Waals and Coulomb interactions
and forces beyond a cut-off distance of $12$~\AA~ at each time step. 
The interactions between ions were modeled with the Born-Mayer-Huggins 
ion-rigid potential described in work~[\onlinecite{cazorla13}]. The suitability of this
approach for studying the energy, structural, and superionic properties of CaF$_{2}$ 
at normal pressure has already been demonstrated.~\cite{cazorla13,cazorla14}
We simulated the effect of applying an uniform external electric field by adding a 
force equal to $-q{\bf E}$ on each ion (where $q$ is the corresponding charge).
Following previous works,~\cite{cazorla13,cazorla14,lindan93} we identified the onset of 
superionicity with the appearance of a non-zero slope in the mean squared displacement 
function (MSD) calculated for the fluorine ions. The MSD function is defined as 
$\langle \Delta r^{2}(t) \rangle = \langle \left( r_{i}(t+t_{0}) - r_{i}(t_{0}) \right)^{2} \rangle$~,  
where $r_{i}(t)$ is the position of atom $i$ at time $t$, $t_{0}$ an arbitrary time
origin, and $\langle \cdots \rangle$ denotes average over F$^{-}$ ions and time origins. 

\begin{figure}[t]
\centerline
        {\includegraphics[width=1.0\linewidth]{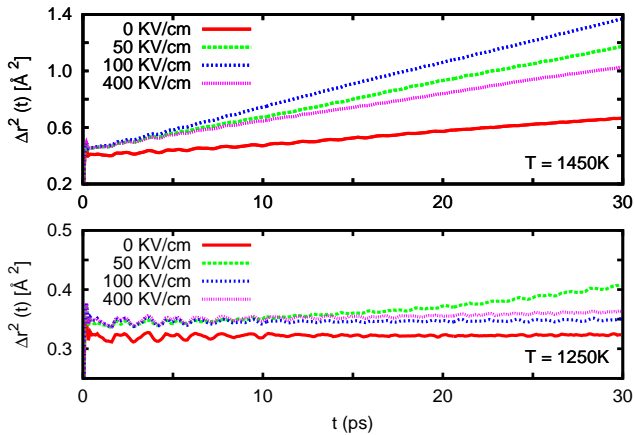}}
\vspace{-0.10cm}
\caption{Anionic MSD function calculated in CaF$_{2}$ considering different temperatures
         and values of the electric field.}
\label{fig:results}
\end{figure}

$\emph{Results.}$~Figure~\ref{fig:results} shows the fluorine MSD function calculated 
at $T<T_{s}$ and $T>T_{s}$ [where $T_{s} = 1350~(25)$~K is the transition temperature 
computed at $E = 0$]~\cite{cazorla13} considering different values of the electric 
field. $E$ was applied along the three inequivalent crystalline directions $[1 0 0]$, 
$[1 1 0]$, $[1 1 1]$, and in all the cases we obtained equivalent results. In our 
analysis we only consider electric fields smaller than $500$~KV~cm$^{-1}$ since otherwise we
found that the Ca$^{2+}$ cations started to drift. At $T = 1250$~K, that is, about $100$~K 
below $T_{s}$, the onset of superionicity appears at $E = 50$~KV~cm$^{-1}$. When the 
module of the electric field is further increased, however, ionic conductivity disappears. 
Such a counterintuitive effect can be understood as follows. At small values of $E$, the 
premature creation of Frenkel defects (i.e., simultaneous formation of an interstitial-vacancy
pair) is estimulated by the presence of the external electric 
field and the accompanying increase in the entropy, which minimises the free energy of the 
sytem. Importantly, at these conditions the diffusion of the F$^{-}$ ions is Brownian 
(that is, $\langle r(t) \rangle = 0$), exactly as it is observed in the crystal at $T > T_{s}$ 
in the absence of an electric bias. Eventually, as $E$ is increased, the diffusion of 
the anions starts being non-erratic ($\langle r(t) \rangle \neq 0$) and thus the entropy 
of the system decreases and the Coulomb interactions between ions are not optimal. 
Consequently, the anions prefer to remain in the ordered normal state, in which the total 
electrostatic energy is most favourable, in order to lower the free energy of the crystal. 
At $T = 1450$~K,  superionicity is fully developed and the effect of applying an electric 
field is to further promote the diffusion of anions. The ionic conductivity is maximum at 
$E = 100$~KV~cm$^{-1}$, however under increasing electric bias the slope of the MSD function 
is reduced (although this is always positive and larger than the obtained at $E = 0$). 
The origins of this effect can be understood in terms of similar entropy and electrostatic 
energy arguments than explained above. Finally, we calculated the heat capacity and change 
of entropy associated to the superionic transition in CaF$_{2}$ at $E \le 100$~KV~cm$^{-1}$ 
and $T \le 1350$~K. We found that the resulting change of temperature assuming 
adiabatic conditions generally is $|\Delta T / T| \approx 1$~\%. 

$\emph{Summary.}$~In conclusion, our molecular dynamics work shows that is possible to 
vary $T_{s}$ in FIC by applying relatively small external electrical fields. This effect 
has the potential to change our paradigm in the search of new electrocaloric materials, 
which are promising for solid-state cooling applications. Analogous studies performed on 
similar FIC with lower $T_{s}$'s (e.g., $\alpha$-PbF$_{2}$) are highly desirable.   

$\emph{Acknowledgments.}$~This research was supported under the Australian 
Research Council's Future Fellowship funding scheme (projects number RG134363 and RG151175).

\end{document}